\documentclass[pra,aps,twocolumn,superscriptaddress,showpacs,reprint]{revtex4}
\usepackage{amsfonts}
\usepackage{amsmath}
\usepackage{amssymb}
\usepackage{graphicx}
\usepackage{epstopdf}

\providecommand{\U}[1]{\protect\rule{.1in}{.1in}}

\begin{document}

\preprint{HEP/123-qed}
\title[ ]{Probing quantum gravity effects with ion trap}
\author{Yue-Yue Chen}
\affiliation{State Key Laboratory of High Field Laser Physics, Shanghai Institution of
Optics and Fine Mechanics, Chinese Academy of Sciences, Shanghai 201800,
China}
\author{Xun-Li Feng}
\email{xlfeng@siom.ac.cn}
\affiliation{State Key Laboratory of High Field Laser Physics, Shanghai Institution of
Optics and Fine Mechanics, Chinese Academy of Sciences, Shanghai 201800,
China}
\affiliation{Centre for Quantum Technologies, National University of Singapore, 2 Science
Drive 3, Singapore 117542}
\author{C. H. Oh}
\affiliation{Centre for Quantum Technologies, National University of Singapore, 2 Science
Drive 3, Singapore 117542}
\author{Zhi-Zhan Xu}
\email{zzxu@mail.shcnc.ac.cn}
\affiliation{State Key Laboratory of High Field Laser Physics, Shanghai Institution of
Optics and Fine Mechanics, Chinese Academy of Sciences, Shanghai 201800,
China}

\begin{abstract}
The existence of minimal length scale has motivated the proposal of
generalized uncertainty principle, which provides a potential routine to
probe quantum gravitational effects in low-energy quantum mechanics
experiment. Hitherto, the tabletop experiment of testing deviations from
ordinary quantum mechanics are mostly based on microscopic objects. However,
the feasibility of these studies are challenged by the recent study of
spacetime quantization for composite macroscopic body. In this paper, we
propose a scheme to probe quantum gravity effects by revealing the
deviations from predictions of Heisenberg uncertainty principle. Our scheme
focus on manipulating the interaction sequences between external laser
fields and a single trapped ion to seek evidence of spacetime quantization,
therefore reduce the complicity induced by large bodies to some extent. The
relevant study for microscopic particles is crucial considering the lack of
satisfactory theories regarding basic properties for multi-particles in the
framework of quantum gravity. Meanwhile, we are managed to set a new upper
limit for deformation parameter.
\end{abstract}

\volumeyear{year}
\volumenumber{number}
\issuenumber{number}
\eid{identifier}
\date[Date text]{date}
\received[Received text]{date}
\revised[Revised text]{date}
\accepted[Accepted text]{date}
\published[Published text]{date}
\startpage{1}
\endpage{102}
\maketitle

\section{Introduction}

Quantum gravity is referred to a theory unifying the general relativity and
quantum mechanics. The primary obstacle in developing such a theory is
lacking testable experiments of quantum gravitational effects. Previously
studies are usually based on high-energy astronomical events \cite{A,U,F}
with energy in the order of $E_{p}=c\hbar /L_{p}=1.2\times 10^{19}$ GeV,
where the general relativity are expected to merge with quantum physics.
While the emergence of a minimal length scale predicted by various
approaches to quantum gravity provides possibilities to find first ever
experimental evidence in low-energy quantum mechanics realm. Specifically,
the existence of minimal length scale is against the Heisenberg uncertainty
relation and motivates the proposal of a generalized uncertainty principle
(GUP). Thus, it's generally believed that quantum gravity can be tested to
perform high-sensitivity measurement of the uncertainty relation. In this
sense, many proposals are aimed to disclose derivations from the
predictions of ordinary quantum mechanics (QM) based on uncertainty relation.
This motivated a growing number of approaches to search for evidence of
Planck-scale physics which raised the hope to get experimental direct access
to the gravity induced effects. 

Refs.\cite{J} expounded the feasibility of
study Planck-scale physics in a tabletop experiment by observing the motion
of a dielectric macroscopic block through a distance of the order of
Planck's length. Refs. \cite{I} proposed schemes to measure possible
Planck-scale deformation with optomechanics in an unprecedented sensitivity.
Ref. \cite{M} measured the change in the oscillator ground-state energy
induced by modified commutator with gravitational bar detectors. All those
proposals are based on the consensus that the quantization of spacetime for
macroscopic objects is same as that for its constituent particles. However,
Camelia challenged this assertion and point out that the spacetime
quantization is much weaker for center-of-mass of a macroscopic body
compared with fundamental particles constitute it \cite{Camelia}. According to the
conclusion of Camelia, new approaches pertaining fundamental particles
should be came up with to detect quantum gravity effects, instead of
focusing on schemes concerning the center-of-mass motions of macroscopic
objects. 

The ion trap system has been widely studied for its many advantages
like long coherent time. In this paper, we propose a scheme to detect the
quantum gravity effects with a two-level ion trapped in a Paul trap. By
manipulating the phase of classical laser addressing the ion, a sequence of
four interactions between ion and laser is designed such that a phase is
accumulated on a specific ground state during the oscillating period of ion
external motion. By repeating the cycle and controlling cycle times, the
phase obtained by standard quantum mechanics is washed out and thus
the deformation related phase corresponding to the derivation result
from quantum gravity effects can be extracted. After a Hadmard transformation with another
auxiliary level, the deformation of phase is mapped into the population
change, which can be detected with a high accuracy. In the case of a null
result of detection, an upper bound for $\beta _{0}$ can be set. In this way, we provide a method
to perfect the quantum gravity theories with empirical feed
back. Moreover, since only a single ion is concerned, our scheme can
avoid the errors referred in \cite{Camelia} introduced by macroscopic probes.

\section{The generalized uncertainty principle}

The Heisenberg uncertainty principle allows localizing a particle sharply at
a point at the expense of the information on the conjugate momentum. 
While when quantum gravity is considered,
uncertainty principle need to be generalized to incorporate the effect of minimal length scale
\cite{GUP}.

\begin{equation}
\Delta \hat{x}\Delta \hat{p}\geq \frac{\hbar }{2}(1+\beta _{0}(\frac{\Delta
\hat{p}}{M_{p}c})^{2}),
\end{equation}%
where $\beta _{0}$ is the deformed parameters that quantifies the
modification strength, $M_{p}$ is the Planck mass and $M_{p}c^{2}$ is Planck
energy $E_{p}$. It has been proved that GUP is equivalent to a modified
canonical commutator in the following form \cite{K},

\begin{equation}
\lbrack \hat{x},\hat{p}]=i\hbar (1+\beta _{0}(\frac{\hat{p}}{M_{p}c})^{2}).
\end{equation}%
Let us define \cite{S}
\begin{equation}
\hat{x}=x,\hat{p}=p(1+\frac{1}{3}\beta p^{3}).
\end{equation}%
The operators with and without a hat represent deformed and standard
operators respectively, and $\beta =\frac{\beta _{0}}{M_{p}c}$. Note that,
the modified Heisenberg algebra (2) is satisfied to order $\beta $, we thus
neglect the terms of higher order throughout the paper.

\begin{figure}[htbp]
\begin{center}
\includegraphics[width=4.5cm,height=5.5cm]{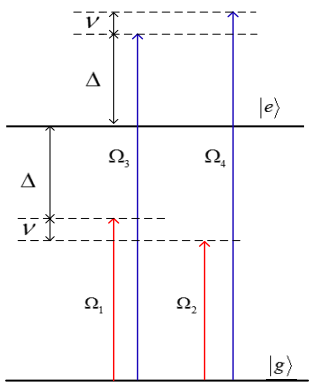}
\end{center}
\caption{(color online) The energy levels of the trapped ion and the
transition driven by four classical lasers. The ion is illuminated by lasers
with opposite detuning.}
\end{figure}
Next, we proceed to elaborate our scheme to measure deformations from ordinary QM.
Considering a two-level ion trapped in a Paul trap, the transition between
ground state $\left\vert g\right\rangle $ and exited state $\left\vert
e\right\rangle $ is driven by four laser fields with frequency $\omega _{i}$
$(i=1,2,3,4),$ as shown in Fig.1. $\omega _{1},\omega _{2}$ are respectively
detuned by $\Delta _{1}$ and $\Delta _{2}$ from the $\left\vert
g\right\rangle \leftrightarrow \left\vert e\right\rangle $ transition and
have a relative detuning $\Delta \omega =\omega _{1}-\omega _{2}=\nu $,
equivalent to the frequency of the vibrational mode of the ion. The
frequencies $\omega _{3},\omega _{4}$ have the opposite detuning
corresponding to $\omega _{1},\omega _{2}$ respectively while the same
relative detuning $\Delta \omega =\omega _{4}-\omega _{3}=\nu .$ The total
Hamiltonian of the system in the framework of GUP takes the form

\begin{eqnarray}
H &=&\hbar \omega _{eg}\left\vert e\right\rangle \left\langle e\right\vert +%
\frac{\hat{p}^{2}}{2m}+\frac{m\omega ^{2}\hat{x}^{2}}{2}  \notag \\
&&+\sum_{m=1}^{4}\frac{\hbar \Omega _{m}}{2}e^{i\omega _{m}t+i%
\overrightarrow{k}_{m}\cdot \overrightarrow{\hat{r}}+i\phi _{m}}\left\vert
e\right\rangle \left\langle g\right\vert +\text{H}.\text{c}. \\
&=&\hbar \omega _{eg}\left\vert e\right\rangle \left\langle e\right\vert +%
\frac{p^{2}}{2m}+\frac{m\omega ^{2}x^{2}}{2}+\frac{\beta p^{4}}{3m}  \notag
\\
&&+\sum_{m=1}^{4}\frac{\hbar \Omega _{m}}{2}e^{i\omega _{m}t+i%
\overrightarrow{k}_{m}\cdot \overrightarrow{\hat{r}}+i\phi _{m}}\left\vert
e\right\rangle \left\langle g\right\vert +\text{H}.\text{c}.
\end{eqnarray}%
where $\Omega _{m}$, $\overrightarrow{k}_{m}$ and $\phi _{m}$ are Rabi
frequency, wave vector and phase of the $m$th laser field, respectively.
For one dimensional case, we project the wave vectors on $\hat{x}$ direction, 
$\overrightarrow{k}_{m}\cdot \overrightarrow{\hat{r}}=k_{xi}%
\hat{x}$. The Hamiltonian in the interaction picture with respect to $%
H_{0}=\hbar \omega _{eg}\left\vert e\right\rangle \left\langle e\right\vert +%
\frac{\hat{p}^{2}}{2m}+\frac{m\omega ^{2}\hat{x}^{2}}{2}$ are rewritten into
\begin{equation}
H=\sum_{m=1}^{4}\frac{\hbar \Omega _{m}}{2}e^{i\Delta _{m}t+ik_{xm}\hat{x}%
(t)+i\phi _{m}}\left\vert e\right\rangle \left\langle g\right\vert +\text{H}.%
\text{c}.\text{,}
\end{equation}%
where $\Delta _{1}=-\Delta _{4}=\Delta ,\Delta _{2}=-\Delta _{3}=\Delta +\nu
,$ and $\hat{x}(t)$ are the position operators of ion at $t.$ Instead of
taking a continuing interaction between laser fields and ion throughout the
whole oscillator period $T$, we turn on the interaction sharply at every
quarter of the period, $t_{i}\nu =\frac{\pi }{2}i$ $(i=0,1,2,3),$ 
for a relatively short time $t_{p}$ ($t_{p}\ll T$). Regardless of 
the external lasers, the evolution of ion is a modified harmonic oscillation following $H_{0}$. 
In this case, the dynamics of the position operators is
obtained by the unitary transformation $e^{\frac{i}{\hbar }H_{0}t}\hat{x}e^{-%
\frac{i}{\hbar }H_{0}t}$ \cite{YY}
\begin{widetext}
\begin{eqnarray}
\hat{x}(t) &=&x(t)  \notag \\
&=&\sqrt{\frac{\hbar }{2m\nu }}(ae^{-i\nu t}+a^{\dagger }e^{i\nu t})+\frac{%
\beta e^{-3i\nu t}}{12}\sqrt{\frac{\hbar ^{3}m\nu }{2}}[-6e^{2i\nu
t}(-1+e^{2i\nu t}+2it\nu )a+  \notag \\
&&12ie^{3i\nu t}a^{\dagger }(e^{i\nu t}t\nu +\sin \nu t)+(2e^{2i\nu
t}-3+e^{4i\nu t})a^{3}-(12ie^{2i\omega t}t\omega +12ie^{3i\nu t}\sin \nu
t)a^{\dagger }a^{2}  \notag \\
&&+(12ie^{4i\nu t}t\nu +12ie^{3i\nu t}\sin \nu t)a^{\dagger 2}a)+(e^{2i\nu
t}+2e^{4i\nu t}-3e^{6i\nu t})a^{\dagger 3}],
\end{eqnarray}%
\end{widetext}
where $a^{\dag }$($a)$ is the canonical creation (annihilation) operator of
vibrational mode. While when $t\in \lbrack t_{i},t_{i}+t_{p}]$, where lasers are turned on for a sufficient short duration $t_{p},$ the harmonic evolution can
be neglected and $\hat{x}(t)=\hat{x}%
(t_{i}).$

In the case of large detuning $\Delta \gg \Omega _{i}$, we may adiabatically
eliminate the excited atomic state $\left\vert e\right\rangle $ since no
population transfers to this state providing the ion is initially populated
on the ground state. 
Thus, with James method \cite{D}, we obtain a
effective Hamiltonian for the interaction between ion and laser fields
during the time interval $[t_{i},t_{i}+t_{p}]$,

\begin{equation}
\begin{split}
H_{eff}=&\widetilde{\Omega }[ (-e^{i(k_{x1}-k_{x2})x(t_{i})+i(\phi _{1}-\phi
_{2})} \\
&+e^{i(k_{x4}-k_{x3})x(t_{i})+i(\phi _{4}-\phi _{3})})e^{-i\nu t}+\text{H.c.}%
] \left\vert g\right\rangle \left\langle g\right\vert .
\end{split}%
\end{equation}%
Where $\widetilde{\Omega }=\frac{\hbar \Omega _{1}\Omega _{2}(\Delta
_{1}+\Delta _{2})}{8\Delta _{1}\Delta _{2}},$ and we have assumed that $%
\Omega _{1}=\Omega _{3},\Omega _{2}=\Omega _{4}$ to eliminate the
time-independent stark shift. Since oscillating frequency $\nu<<\Delta$,
the approximation used for $%
\hat{x}(t)$ is feasible for $H_{eff}$, thus $H_{eff}(t)=H_{eff}(t_{i})$ for $t\in \lbrack t_{i},t_{i}+t_{p}].$

By manipulating the relative phase of lasers, we are able to interchange
the canonical position operator $x$ and momentum $p$ every quarter of
harmonic evolution period. After using four interactions separated by a
quarter period, a phase containing the contribution from GUP is accumulated
on the ground state $\left\vert g\right\rangle $. Specifically, at the
initial time $t_{0},$\ the phases are adjusted to be in the relation $\phi
_{1}-\phi _{2}=\phi _{4}-\phi _{3}=\frac{\pi }{2}.$ Thus the effective
Hamiltonian is simplified to

\begin{equation}
\begin{split}
H_{eff}=&i\widetilde{\Omega }[ (-e^{i(k_{x1}-k_{x2})x(t_{0})} \\
&+e^{i(k_{x4}-k_{x3})x(t_{0})})e^{-i\nu t_{0}}+\text{H.c.}] \left\vert
g\right\rangle \left\langle g\right\vert .
\end{split}%
\end{equation}%
In the Lamb-Dicke regime, the interaction Hamiltonian takes the form

\begin{eqnarray}
H_{eff} &=&\widetilde{\Omega }[ (-(1-i(k_{x1}-k_{x2})x(t_{0}))  \notag \\
&&+1+i(k_{x4}-k_{x3})x(t_{0}))e^{-i\nu t_{0}}+\text{H.c.}] \left\vert
g\right\rangle \left\langle g\right\vert  \notag \\
&=&-2\widetilde{\Omega }\Delta kx(t_{0})\cos \nu t_{0}\left\vert
g\right\rangle \left\langle g\right\vert ,
\end{eqnarray}%
where $\Delta k=k_{x4}+k_{x2}-k_{x3}-k_{x1}$. Substitute the $t_{0}=0$ and $%
x(t_{0})$ obtained from Eq(7) into Eq.(10), we can get the time-independent
Hamiltonian

\begin{equation}
\begin{split}
H_{eff}=&-\sqrt{\frac{2\hbar }{m\nu }}\widetilde{\Omega }\Delta k(a^{\dag
}+a)\left\vert g\right\rangle \left\langle g\right\vert  \notag \\
=&-2\widetilde{\Omega }\Delta kx(0)\left\vert g\right\rangle \left\langle
g\right\vert ,
\end{split}%
\end{equation}%
where $x(0)=\sqrt{\frac{\hbar }{2m\nu }}(a^{\dag }+a)$ stands for the
canonical position operator in Sch\"{o}dinger picture. Considering $\Delta
_{1},\Delta _{2}\gg \nu ,$ $\Delta _{1}\approx \Delta _{2}\approx \Delta ,$
the time evolution operator $U_{0}(t)$ takes the form

\begin{equation}
U_{0}(t)=e^{i\eta x(0)\left\vert g\right\rangle \left\langle g\right\vert
t},t\in \lbrack 0,t_{p}],
\end{equation}%
where $\eta =\frac{\Delta k\Omega _{1}\Omega _{2}}{2\Delta }.$ For $t_{0}+t_{p}<t<t_{1},$ the
interaction between laser and ion is turned off and the
external motion of ion is just a modified harmonic oscillation following $H_{0}$.
At $t_{1}=T/4$, $\hat{x}(0)$ evolves to $\hat{x}(t_{1})$.
Meanwhile we change the phases of laser fields to $%
\phi _{1}=\phi _{2},\phi _{4}=\phi _{3}.$ Thus the corresponding Hamiltonian
for the second time interval $t\in \lbrack \frac{T}{4},\frac{T}{4}+t_{p}]$
takes the form (see Appendix B for more details),

\begin{eqnarray}
H_{eff} &=&2\widetilde{\Omega }\Delta kx(t_{1})\sin \nu t_{1}\left\vert
g\right\rangle \left\langle g\right\vert ,
\end{eqnarray}
and time evolution operator
\begin{eqnarray}
U_{1}(t) &=&e^{-i\frac{\eta }{m\nu }p(0)\left\vert g\right\rangle
\left\langle g\right\vert t}e^{i\beta \xi t^{4}\left\vert g\right\rangle
\left\langle g\right\vert },
\end{eqnarray}%
where $\xi =\frac{\beta \hbar ^{3}\pi }{256m\nu }(\frac{\Delta k\Omega
_{1}\Omega _{2}}{\Delta })^{4}$. Similarly, we adjust the phases at $t_{2}=T/2$
to satisfy $\phi _{1}-\phi _{2}=\phi _{4}-\phi _{3}=-\frac{\pi }{2}.$ Thus,
the interaction Hamiltonian during $[\frac{T}{2},\frac{T}{2}+t_{p}]$ takes
the form
\begin{eqnarray}
H_{eff} &=&2\widetilde{\Omega }\Delta kx(t_{2})\cos \nu t_{2}\left\vert
g\right\rangle \left\langle g\right\vert ,
\end{eqnarray}
subsequently,
\begin{eqnarray}
U_{2}(t) &=&e^{-i\eta x(0)\left\vert g\right\rangle \left\langle
g\right\vert t}e^{i2\beta \xi t^{4}\left\vert g\right\rangle \left\langle
g\right\vert }.
\end{eqnarray}%
After another quarter of vibrational period, the laser phases
are adjusted to $\phi _{1}-\phi _{2}=\phi _{4}-\phi _{3}=\pi .$ By this
time, the interaction Hamiltonian $H_{eff}$ and $U_{3}(t)$ during $[\frac{3T%
}{4},\frac{3T}{4}+t_{p}]$ are
\begin{eqnarray}
H_{eff} &=&-2\widetilde{\Omega }\Delta kx(t_{3})\sin \nu t_{3}\left\vert
g\right\rangle \left\langle g\right\vert ,  \notag \\
U_{3}(t) &=&e^{^{i\frac{\eta }{m\nu }p(0)\left\vert g\right\rangle
\left\langle g\right\vert t}}e^{i3\beta \xi t^{4}\left\vert g\right\rangle
\left\langle g\right\vert }.
\end{eqnarray}%
Eventually, assuming the ion is initially populated on the ground state $%
\left\vert g\right\rangle ,$ the final state of the system after a round
trip consisted of four interaction sequences is

\begin{eqnarray}
\Psi (T) &=&U_{3}(t_{p})U_{2}(t_{p})U_{1}(t_{p})U_{0}(t_{p})\left\vert
g\right\rangle  \notag \\
&=&e^{-i\frac{\hbar }{4m\nu }(\frac{t_{p}\Delta k\Omega _{1}\Omega _{2}}{%
\Delta })^{2}+i\frac{3\beta \hbar ^{3}\pi }{128m\nu }(\frac{t_{p}\Delta
k\Omega _{1}\Omega _{2}}{\Delta })^{4}}\left\vert g\right\rangle.
\end{eqnarray}%
Conspicuously, an additional phase
proportional to $\beta $ is produced by the deformation of the canonical
commutator due to the existence of minimal length scale. Particularly, by
choosing the parameters properly such that $\frac{\hbar }{4m\nu }(\frac{%
t_{p}\Delta k\Omega _{1}\Omega _{2}}{\Delta })^{4}=2\pi m,$ $m$ is an
integer, the contribution from the $\beta $ term only can be extracted. In this
case, the deformations of the ordinary quantum mechanics are present in a form of
accumulated phase during the periodic evolution, which can be measured 
straightforwardly.

To enlarge the effects induced by quantum gravity, we repeat the procedure
for another $N-1$ times. Specifically, we repeat the interaction sequences
subsequently at the time $t_{i}\nu =\frac{\pi }{2}i$ $(i=0,1,2,...4N-1).$
Note that, the $\beta $ terms actually form a arithmetic progression with a tolerance $d=i\frac{\beta
\hbar ^{3}\pi }{256m\nu }(\frac{t\Delta k\Omega _{1}\Omega _{2}}{\Delta }%
)^{4}$ while the ordinary phase remains unchanged for
every cycle. In this way, the final state at $t_{f}$ after $N$ times cycles
can be calculated easily
\begin{equation}
\Psi (t_{f})=e^{i\phi }\left\vert g\right\rangle =e^{i(\phi _{0}+\delta \phi
)}\left\vert g\right\rangle ,
\end{equation}%
where%
\begin{eqnarray}
\phi _{0} &=&-\frac{N\hbar }{4m\nu }(\frac{t_{p}\Delta k\Omega _{1}\Omega
_{2}}{\Delta })^{2}, \\
\delta \phi &=&(4N-1)2N\frac{\beta \hbar ^{3}\pi }{256m\nu }(\frac{%
t_{p}\Delta k\Omega _{1}\Omega _{2}}{\Delta })^{4}
\end{eqnarray}%
Note that, $\phi _{0}$ is corresponding to the phase governed by standard
quantum mechanics, while $\delta \phi $ is a possible deviation result from
GUP.

\section{Measurement of the deformation}

\bigskip Now we proceed to apply our theory to a real system and propose a
scheme to measure the phase. $^{171}$Yb$^{+}$ ion as a popular element 
widely used in ion trap system has been a candidate for studies of
interactions with ultracold atoms and quantum information processing \cite%
{Yb}. Recently it has found application in fluorescence detection with high
speed and high fidelity \cite{FD1,FD2}. We use $^{2}P_{1/2}$ as the excited
state $\left\vert e\right\rangle $ and $^{2}S_{1/2}$ as the ground state $%
\left\vert g\right\rangle .$ To probe the deformation related phase, we need
to take another ancillary state $^{2}D_{3/2}$ (denoted by $\left\vert
r\right\rangle )$\ into consideration. The lifetime of $\left\vert
e\right\rangle $ is not relevant in our scheme due to the adiabatic
elimination adopted above and the lifetime of ground states $\left\vert
g\right\rangle $ is considered infinite. The lifetime of metastable $%
^{2}D_{3/2}$ state is $52$ ms \cite{C} which is three orders of magnitude
larger than the time scale required for fluorescence detection \cite{FD2}
and there is no population in $^{2}D_{3/2}$ before the Hadmard
transformation. Therefore the lifetime of $^{2}D_{3/2}$ can be ignored. Thus
the number of loops $N$ is only limited by the storage time of the ion trap,
which is at least several hours for a $^{171}$Yb$^{+}$ ion. The parameters
are chosen based on experimental works \cite{FD1} to meet the adopted approximations: $M=173.04$ u, $\nu
=0.18\times 2\pi $ MHz, $t_{p}=0.56$ $\mu $s, $\Omega _{1}=\Omega _{2}=2$
GMz, $\Delta =12$ GHz, $\left\vert \overrightarrow{k}_{i}\right\vert =\frac{%
2\pi }{\lambda }=2.7\times 2\pi $ rad/$\mu $m, $\Delta k=1.54\left\vert
\overrightarrow{k}_{i}\right\vert $. 
With $N=1.944\times 10^{9}$ ($t$ $\sim$ 3 hours) and $%
\beta _{0}$ $\sim$ $10^{33}$ \cite{S}, 
the total phase accumulated is $\phi
=-0.1167241\pi $ and the deformation part $\delta \phi =0.293155\pi $. 

To read out the phase on $\left\vert g\right\rangle ,$ we initially prepare
the ion in state $\Psi (0)=\frac{1}{\sqrt{2}}(\left\vert r\right\rangle
+\left\vert g\right\rangle ).$ Without the effects induced by quantum
gravity, the final state $\Psi (t_{f})$ will be identical with $\Psi (0)$
and $\Psi (t_{f})\rightarrow \left\vert g\right\rangle $ after a Hadamard
transformation. While when gravity effects are considered, $\Psi (t_{f})=%
\frac{1}{\sqrt{2}}(\left\vert r\right\rangle +e^{i\phi }\left\vert
g\right\rangle )$. After a Hadamard transformation, $\Psi (t_{f})\rightarrow
e^{i\frac{\phi }{2}}(\cos \frac{\phi }{2}\left\vert g\right\rangle +i\sin
\frac{\phi }{2}\left\vert r\right\rangle )$ and the population on $%
\left\vert r\right\rangle $ is $P_{r}=\left\vert \sin \frac{\phi }{2}%
\right\vert ^{2}.$ As a result, the population difference between
measurement and standard results $\delta P_{r}=\left\vert \sin \frac{\phi }{2%
}\right\vert ^{2}-$ $\left\vert \sin \frac{\phi _{0}}{2}\right\vert ^{2}$
can signal the effects induced by quantum gravity. On the other hand, the
null results of precision measurement may predict an upper bound for $\beta
_{0}$, which is the case $\delta P_{r}$ is below measurement accuracy. A
commonly used method for estimating the population is based on accurately
measuring the fluorescence and excited-state fraction (ESF) in the MOT \cite%
{R}. According to \cite{MOT}, the present experimental setup used by
Flechard's group has a sensitivity better than $10^{-3}$ for a Rb target.
\cite{RG} proposed a novel technique to measure the branching fractions of $%
^{40}$Ca$^{+}$ based on repetitive optical pumping, which improved the
accuracy of precision measurement to about 1 part in $10^{5}$. With the
state of the art accuracy, we are able to set a new bound $\beta
_{0}<10^{24} $, which would improve the existing bounds for $\beta _{0}$ by
nine orders of magnitude.
\begin{table*}[tbp]
\caption{The primary parameters, energy levels \protect\cite{Be} and upper
bounds given by three different ion species. The accuracy of precision
measurement puts a straightforward upper limit to the population derivation
from standard quantum mechanics $\protect\delta P_{r}<10^{-5}$. Other
parameters beyond the list are same as those in text. }\centering%
\begin{tabular}{ccccccccc}
\hline\hline
Species & $\lambda $ (nm) & $N$ $(10^{9})$ & $\nu /2\pi $ (KHz) & $\Delta $k/%
$\left\vert k\right\vert $ (rad) & $\left\vert e\right\rangle $ & $%
\left\vert g\right\rangle $ & $\left\vert r\right\rangle $ & $\beta _{0}$ \\
\hline
$^{171}$Yb$^{+}$ & 369.5 & 1.944 & 180 & 1.54 & $^{2}P_{1/2}$ & $^{2}S_{1/2}$
& $^{2}D_{3/2}$ & $10^{24}$ \\
$^{40}$Ca$^{+}$ & 393 & 5.4 & 500 & 1.31 & $^{2}P_{3/2}$ & $^{2}S_{1/2}$ & $%
^{2}D_{5/2}$ & $10^{25}$ \\
$^{9}$Be$^{+}$ & 313 & 1 & 0.07 & 0.01 & $^{2}P_{3/2}(F=2)$ & $%
^{2}S_{1/2}(F=2)$ & $^{2}S_{1/2}(F=1)$ & $10^{18}$ \\ \hline\hline
\end{tabular}%
\end{table*}
Table 1 compares the parameters and the corresponding upper bounds for $%
^{171}$Yb$^{+},$ $^{40}$Ca$^{+}$ and $^{9}$Be$^{+}$, the species generally
used in ion trap system. From the table we can see, with an increasing
storage time, lower vibrational frequency and a more accurate measurements
in the future, the upper bounds are expected to be tightened by several
orders of magnitude. Note that the transition $\left\vert g\right\rangle
\leftrightarrow \left\vert e\right\rangle $ of $^{9}$Be$^{+}$ can be
selected by lasers with $\sigma ^{+}/\sigma ^{-}$ polarization to avoid
activation of $\left\vert r\right\rangle $. Interestingly, $\phi _{0}=2\pi m$
for the case of $^{9}$Be$^{+}$ with the parameters listed. Thus, the phase
corresponding to the standard quantum mechanics is eliminated, and the phase
accumulated after $10^{9}$ loops is only attribute to quantum gravity.

\section{Discussion}

So far, our scheme is based on the assumption that the interaction with
environment can be neglected. Actually, decoherence effect such as thermal
motion of ion is not likely to spoil the fidelity due to the virtual
excitations of vibrational mode. The creation and annihilation operators of
vibrational mode are disappeared after the four interaction sequences,
leaving the vibrational mode invariable. The independence of vibrational
mode remind us of the elimination of SM model \cite{SM}, while our scheme is
realized on a different mechanism based on manipulating of laser phases. The
key of our scheme is the precision control of the interaction time $t_{p}$
such that the dynamics of frequency down to $\nu $ scale can be neglected
during interaction time interval $t_{p}$. To do this the external laser
fields are required to turn on and off within a few ps and the trap
frequency is in KHz scale. The bounds set by our scheme can be tightened
with a lower trap frequency, more accurate measurements and longer trap
lifetime. The basic properties of macroscopic bodies, such as spacetime
geometry and measurement process, are not available at the moment, and
microscope atoms are much more likely to be affected by the full strength of
Planck-scale effects than macroscopic reality. Thus, our scheme provides a
method to detect possible effects induced by quantum gravity and circumvents
the unpredictable deformation of spacetime quantization when probing with
microscopic body. At the same time, the null results of probing can be used
to explore the bounds of quantum gravity parameters and signal a
intermediate length scale smaller than Planck scale.

\section*{Acknowledgements}

Enlightened discussions with Prof. Yu Sixia are gratefully acknowledged.
This work was supported by NSFC (Grant Nos 11074079 and 11174081) and the
National Research Foundation and Ministry of Education, Singapore (Grant No.
WBS: R-710-000-008-271). \setcounter{equation}{0} \renewcommand{%
\theequation}{A\arabic{equation}}

\section*{Appendix}

The calculation for $U_{0}(t)$ is straightforward since no $\beta $ terms
are involved while the calculation for $U_{2}(t)$ and $U_{3}(t)$ are
basically the same with $U_{1}(t)$. Thus we take the interaction between
laser and ion during $[\frac{T}{4},\frac{T}{4}+t_{p}]$ as an example to give
the detailed derivation for time evolution operator$.$ By substitute the
expression of $x(t_{1})$ into Eq.(11), we get

\begin{equation}
\begin{split}
H_{eff}=& (-i\sqrt{\frac{2\hbar }{m\nu }}\widetilde{\Omega }\Delta
k(a-a^{\dag })) \\
& -\frac{1}{3}\hbar \widetilde{\Omega }\Delta k\beta \sqrt{\frac{\hbar m\nu
}{2}}(2i(a^{\dag }-a)^{3}+\pi (a^{\dag }+a)^{3} \\
& -(4i+\pi )a^{\dag 3}+(4i-\pi )a^{3}).
\end{split}%
\end{equation}%
Subsequently, the time evolution operator takes the form
\begin{eqnarray}
U_{1}(t) &=&\exp [A+B],  \notag \\
A &=&-t\sqrt{\frac{\hbar }{2m\nu }}\frac{\Delta k\Omega _{1}\Omega _{2}}{%
2\Delta }(a-a^{\dag }),  \notag \\
B &=&\frac{\hbar \Delta k\beta t\Omega _{1}\Omega _{2}}{12\Delta }\sqrt{%
\frac{\hbar m\nu }{2}}(2(a-a^{\dag })^{3}+  \notag \\
&&i\pi (a^{\dag }+a)^{3}+(4-i\pi )a^{\dag 3}-(4+i\pi )a^{3}).
\end{eqnarray}%
To simplify the four interaction sequences $%
U=U_{0}(t_{p})U_{1}(t_{p})U_{2}(t_{p})U_{3}(t_{p})$ after a round, we
separate the items with and without $\beta $ in each $U_{i}(t)$ with
Zassenhaus formula

\begin{equation}
\begin{split}
& \exp (A+B)=\exp (A)\exp (B)\prod\limits_{i=1}^{\infty }\exp (C_{i}), \\
& C_{1}=-[A,B]/2, \\
& C_{2}=[A,[A,B]]/6+[B,[A,B]]/3, \\
& C_{3}=-([B,[A,[A,B]]]+[B,[B,[A,B]]])/8-[A,[A,[A,B]]]/24.
\end{split}%
\end{equation}
$C_{i},i>3$ are functions of higher nested commutators. Substitute Eq.(A2)
into Eq(A3),

\begin{eqnarray}
C_{1} &=&\frac{i\beta }{32}(\frac{\hbar t\Delta k\Omega _{1}\Omega _{2}}{%
\Delta })^{2}(2\pi   \notag \\
&&+(4i+\pi )a^{2}+4\pi a^{\dag }a+(-4i+\pi )a^{\dag 2}),  \notag
\end{eqnarray}%
\begin{eqnarray}
C_{2} &=&\frac{i\beta }{96}(\frac{\hbar t\Delta k\Omega _{1}\Omega _{2}}{%
\Delta })^{3}\sqrt{\frac{1}{2\hbar m\nu }}((4i+3\pi )a  \notag \\
&&+(-4i+3\pi )a^{\dag }),  \notag \\
C_{3} &=&\frac{i\beta \pi }{256\hbar m\nu }(\frac{\hbar t\Delta k\Omega
_{1}\Omega _{2}}{\Delta })^{4},  \notag \\
C_{i} &=&0\text{ }(i\geq 4).
\end{eqnarray}
The parameters are chosen properly such that $\frac{t\Delta k\Omega
_{1}\Omega _{2}}{\Delta }\sqrt{\frac{\hbar }{2m\nu }}\gg 1$ is satisfied. In
that case, for the items with $\beta $, only the leading order in $\frac{%
t\Delta k\Omega _{1}\Omega _{2}}{\Delta }\sqrt{\frac{\hbar }{2m\nu }}$ is
relevant and thus is saved in Eq. (13) besides the item without $\beta $.


\begin{thebibliography}{99}
\bibitem{A} {A. Camelia, G. Ellis, J. Mavromatos, N. E. Nanopoulos, D. V.
Sarkar and S, Nature \textbf{393}, 763 (1998).}

\bibitem{U} {U. Jacob and T. Piran, Nat. Phys. \textbf{7}, 87 (2007).}

\bibitem{F} {F. Tamburini, C. Cuofano, M.D. Valle and R. Gilmozzi, Astron.
Astrophys. \textbf{533}, 1 (2011).}

\bibitem{J} {J. D. Bekenstein, Phys. Rev. D \textbf{86}, 124040 (2012).}

\bibitem{I} {I. Pikovski, M. R. Vanner, M. Aspelmeyer, M. S. Kim, and C.
Brukner, Nat. Phys. \textbf{8}, 393 (2012).}

\bibitem{M} {F. Marinet al., Nat. Phys. \textbf{9}, 71 (2013).}
\bibitem{Camelia}{G. Amelino-Camelia, Phys. Rev. Lett. \textbf{111}, 101301 (2013).}

\bibitem{GUP} {D. Amati, M. Ciafaloni, and G. Veneziano, Phys. Lett. B
\textbf{216}, 41 (1989); M. Maggiore, Phys. Lett. B \textbf{304}, 65 (1993);
L. J. Garay, Int. J. Mod. Phys. A \textbf{10}, 145 (1995)}

\bibitem{K} {A. Kempf, G. Mangano, and R. B. Mann, Phys. Rev. D \textbf{52},
1108 (1995).}

\bibitem{S} {S. Das, Phys. Rev. Lett. \textbf{101}, 221301 (2008).}

\bibitem{YY} {Y.-Y. Chen, X.-L. Feng, C. H. Oh, and Z.-Z. Xu, Phys. Rev. D.}

\bibitem{D} {D. F. James and J. Jerke, Can. J. Phys. \textbf{85}, 625 (2007).%
}

\bibitem{Yb} {J. I. Cirac and P. Zoller, Phys. Rev. Lett. \textbf{74}, 4091
(1995); S. Olmschenk, K. C. Younge, D. L. Moehring, D. N. Matsukevich, P.
Maunz, and C. Monroe, Phys. Rev. A \textbf{76}, 052314 (2007); A. T. Grier,
M. Cetina, F. Oru\v{c}evi\'{c}, and V. Vuleti\'{c}, Phys. Rev. Lett. \textbf{%
102}, 223201 (2009); C. Zipkes, S. Palzer, C. Sias, and M. Kohl, Nature
(London) \textbf{464}, 388 (2010).}

\bibitem{FD1} {S. Ejtemaee, R. Thomas, and P. C. Haljan, Phys.Rev.A \textbf{%
82}, 063419 (2010).}

\bibitem{FD2} {R. Noek, G. Vrijsen, D. Gaultney, E. Mount, T. Kim, P. Maunz,
and J. Kim, Opt. Lett. \textbf{38}, 4735 (2013).}

\bibitem{C} {C. Gerz, J. Roths, F. Vedel, and G. Werth, Z. Phys. D \textbf{8}%
, 235 (1987).}

\bibitem{R} {R. D. Glover, J. E. Calvert, and R. T. Sang, Phys. Rev. A
\textbf{87}, 023415 (2013).}

\bibitem{MOT} {X. Flechard, H. Nguyen, R. Bredy, S. R. Lundeen, M. Stauffer,
H. A. Camp, C. W. Fehrenbach, and B. D. DePaola, Phys. Rev. Lett. \textbf{24}%
, 243005(2003); A. Leredde, A. Cassimi, X. Flechard, D. Hennecart, H. Jouin,
and B. Pons, Phy. Rev. A \textbf{85}, 032710 (2012).}

\bibitem{RG} {R. Gerritsma, G. Kirchmair, F. Z%
\"{}%
ahringer, J. Benhelm, R. Blatt, and C.F. Roos, Eur. Phys. J. D \textbf{50},
13 (2008).}

\bibitem{SM} {A. S}\O {rensen and K. M\O lmer, Phy. Rev. Lett. \textbf{82},
1971 (1999).}

\bibitem{Be} {D. Leibfried, B. DeMarco, V. Meyer, D. Lucas, M. Barrett, J.
Britton, W. M. Itano, B. Jelenkovic acute, C. Langer, T. Rosenband and D. J.
Wineland, Nature, \textbf{422} 412(2003); C. Monroe, D. M. Meekhof, B. E.
King, D. J. Wineland, Science \textbf{272} 1131(1996); R. Blatt, H. Haffner,
C. F. Roos, C. Becher, and F. Schmidt-Kaler, Quantum Inf. Process. \textbf{3}%
, 61 (2004).}
\end{thebibliography}
\end{document}